\documentclass[aps,prb,twocolumn,superscriptaddress,showpacs]{revtex4}

\usepackage{amsmath}
\usepackage{graphicx}

\def\v{\vec}

\def\t{\underline}

\usepackage{color}
\usepackage[normalem]{ulem}

\begin{document}


\title{Screened empirical bond-order potentials for Si-C}

\author{Lars Pastewka}
\affiliation{Department of Physics and Astronomy, Johns Hopkins University,
  3400 North Charles Street, Baltimore, MD 21218, USA}
\affiliation{Fraunhofer-Institut f\"ur Werkstoffmechanik IWM, W\"ohlerstra\ss e 11, 79108 Freiburg, Germany}
\author{Andreas Klemenz}
\affiliation{Fraunhofer-Institut f\"ur Werkstoffmechanik IWM, W\"ohlerstra\ss e 11, 79108 Freiburg, Germany}
\author{Peter Gumbsch}
\affiliation{Fraunhofer-Institut f\"ur Werkstoffmechanik IWM, W\"ohlerstra\ss e 11, 79108 Freiburg, Germany}
\affiliation{Karlsruher Institut f\"ur Technologie, IAM-ZBS, Kaiserstra\ss e 12, 76131 Karlsruhe, Germany}
\author{Michael Moseler}
\affiliation{Fraunhofer-Institut f\"ur Werkstoffmechanik IWM, W\"ohlerstra\ss e 11, 79108 Freiburg, Germany}
\affiliation{Universit\"at Freiburg, Physikalisches Institut, Hermann-Herder-Stra\ss e 3, 79104 Freiburg, Germany}
\affiliation{Freiburger Materialforschungszentrum, Stefan-Meier-Stra\ss e 21, 79104 Freiburg, Germany}

\date{\today}

\begin{abstract}
Typical empirical bond-order potentials are short ranged and give ductile instead of brittle behavior for materials such as crystalline silicon or diamond.
Screening functions can be used to increase the range of these potentials.
We outline a general procedure to combine screening functions with bond-order potentials that does not require to refit any of the potential's properties.
We use this approach to modify Tersoff's [Phys. Rev. B 39, 5566 (1989)], Erhart \& Albe's [Phys. Rev. B 71, 35211 (2005)] and Kumagai et al.'s [Comp. Mater. Sci. 39, 457 (2007)] Si, C and Si-C potentials.
The resulting potential formulations correctly reproduce brittle materials response, and give an improved description of amorphous phases.
\end{abstract}


\maketitle

\section{Introduction}

Empirical and semi-empirical classical interatomic potentials have been used in computer simulations for more than two decades.
Bond-order potentials (BOPs)~\cite{Pastewka:2012p493} --- a class of semi-empirical formulations --- have proven to yield reasonably accurate potential energy landscapes for covalently bonded~\cite{
Tersoff:1986p632,
Tersoff:1988p2879,
Tersoff:1988p6991,
Tersoff:1988p9902,
Tersoff:1989p5566,
Brenner:1990p9458,
Brenner:1992p1948,
Albe:1998p111,
Oleinik:1999p8500,
Matsunga:2000p48,
Brenner:2002p783,
Albe:2002p035205,
Ni:2004p7261,
Erhart:2005p035211,
Mrovec:2007p230,
Gillespie:2007p155207,
Shan:2010p235302,Yu:2007p085311,
Knippenberg:2012p164701,
Schall:2012}
and metallic~\cite{
Pettifor:1991p439,
Pettifor:1995p24,
Albe:2002p195124,
Mrovec:2004p094115,
Juslin:2005p123520,
Erhart:2006p6585,
Aoki:2007p154,
Muller:2007p326220,
Mrovec:2007p104119,
Henriksson:2009p144106,
Mrovec:2011p246402}
materials.
The bond-order approach can be systematically derived from the tight-binding approximation.~\cite{
Pettifor:1991p439,
Pettifor:1995p24,
Horsfield:1996p12694,
Oleinik:1999p8500,
Pettifor:1999p8487}
This furnishes the hope that although simple, BOPs should show transferability to a wide number of situations.

The rigorous derivation of BOPs by Pettifor and co-workers~\cite{
Horsfield:1996p12694,
Oleinik:1999p8500,
Pettifor:1999p8487} was predated by empirical formulations that are the scope of this article.~\cite{
Abell:1985p6184,
Tersoff:1986p632,
Tersoff:1988p2879,
Tersoff:1988p6991,
Tersoff:1988p9902,
Tersoff:1989p5566}
The parameters of these empirical BOPs are adjusted to match ground-state properties, such as the cohesive energies or elastic constants, for one or more phases of an element or compound.
For covalently bonded materials, the interaction between atoms is usually limited to nearest-neighbors and also limited to short distances.
Both limitations are independent of each other, although a short interaction range is typically used to limit the interaction to nearest-neighbors.
This is possible because in crystalline structures second-nearest neighbors are well-separated from first neighbors.
They show up as well-distinguishable peaks in the atomic pair distribution functions.
It is less clear that limiting the interaction range works in liquids or amorphous solids where such separation is not necessarily given.
Additionally, the short range makes the description of transition events such as the dissociation of a bond inaccurate.
Qualitatively unphysical behavior is obtained in particular when a transition is driven by external forces.~\cite{
Marder:1999p48,
Pastewka:2008p161402,
Pastewka:2012p493}

An example of this latter problem is a crack that is driven through a
brittle material.
In contrast to physical reality, empirical BOPs consistently predict ductile behavior for materials such as silicon or carbon.~\cite{Marder:1999p48}
This problem is usually circumvented by using full quantum calculations~\cite{Perez:2000p5347,Perez:2000p4517} or by embedding a quantum region around the crack tip in a classical potential.~\cite{Kermode:2008p1224,Moras:2010p075502}
However, the interaction between multiple cracks, driving a crack in an amorphous material, or a series of mode II cracks such as a tribological interface~\cite{
Harrison:1994p10399,
Gerde:2001p285,
Harrison:2008p354009,
Pastewka:2008p1136,
Schall:2010p5321,
Pastewka:2010p49,
Pastewka:2011p34} would be notoriously difficult to model with either approach.
A classical interatomic potential that reproduces brittle fracture is therefor highly desirable.

We have shown in an earlier work~\cite{Pastewka:2008p161402} that brittle behavior can be restored by decoupling the condition for nearest-neighbor relationship from the range of the potential.
This potential was based on the second-generation reactive empirical bond-order potential (REBO2),~\cite{Brenner:2002p783} and nearest-neighbor relationship was determined using the screening functions first introduced by Baskes and co-workers~\cite{Baskes:1994p505} in the context of the modified embedded atom method.
This modification kept the REBO2's ground-state properties of crystalline structures and molecules untouched.
Here, we add two silicon-carbide potentials and one pure silicon potential to the family of screened BOPs.
The first two are based on Tersoff's 3rd~\cite{Tersoff:1989p5566} and Erhart \& Albe's~\cite{Erhart:2005p035211} potentials.
The Si-only potential is based on the parameterization by Kumagai et al. that was optimized for the melting point of silicon.~\cite{Kumagai:2007p457}
The particular form of Kumagai's potential has a desirable feature that Tersoff's and Erhart \& Albe's are lacking.
All potentials are modified in a manner that does not change the properties of crystalline ground states.

\section{Second-moment bond-order potentials}

In bond-order potentials of the Tersoff-Brenner type, the cohesive energy $E$ of a structure is expressed as a sum over bonds.
For each bond the energy has a purely repulsive, $\phi(r)$, and a purely attractive, $\beta(r)$, contribution.
The strength of the attractive contribution is modulated by the bond-order, a quantity that depends on the environment of the bond and that is related to Coulson's bond-order concept.~\cite{Coulson:1939p413}
The particular expression we use here is 
\begin{equation}
  E = \frac{1}{2} \sum\limits_{i<j} S_{ij} \left[
  \phi(r_{ij}) - b_{ij} \beta(r_{ij})
  \right]
  \label{bonden}
\end{equation}
where $\phi(r_{ij})$ and $\beta(r_{ij})$ are pairwise positive functions.
The function $b_{ij}$ is the bond-order.
$S_{ij}$ is a switching function that switches the interaction off under certain conditions and will be described in more detail below.

The expression for the bond-order is
\begin{equation}
  b_{ij} = \left( 1 + \chi_{ij}^\eta \right)^{-\delta}
  \label{bij}
\end{equation}
with
\begin{equation}
  \chi_{ij} = \sum\limits_{k\not=i,j}
  S_{ik}
  h(r_{ij}, r_{ik})
  g(\theta_{ijk})
  \label{chiij}
\end{equation}
where
\begin{equation}
  h(r_{ij}, r_{ik})=\exp\left\{\left[2\mu_{ik} (r_{ij} - r_{ik}) \right]^m\right\}.
  \label{h}
\end{equation}
Here, $g(\theta)$ is some function with angular periodicity and $\eta$, $\delta$, $\mu$ and $m$ are free parameters.
Abell~\cite{Abell:1985p6184} used a Bethe lattice analysis to show that the bond-order should be $b_{ij}\propto\sqrt{Z}$ where $Z$ is the local coordination number.
Hence $\eta=1$ and $\delta=1/2$ is the choice consistent with chemical pseudopotential theory.
The bond-order enables a directional dependence of bonding and hence stabilizes the open cage-like structures (e.g. diamond) that covalently bonded materials form.

In what follows, we will discuss potentials where the pairwise functions $\phi$ and $\beta$ are given by exponentials of the form
\begin{equation}
  \beta(r) = \frac{K D_0}{K-1} \exp\left[-\alpha \sqrt{\frac{2}{K}}(r-r_0)\right]
  \label{beta}
\end{equation}
and
\begin{equation}
  \phi(r) = \frac{D_0}{K-1} \exp\left[-\alpha \sqrt{2K}(r-r_0)\right].
  \label{phi}
\end{equation}
With these choices, the dimer potential energy curve $\beta(r)-\phi(r)$ has its minimum at $r_0$ with energy $D_0$ and curvature $\alpha$.
It has been shown~\cite{Abell:1985p6184} that Eqs.~\eqref{beta} and \eqref{phi} together with Eq.~\eqref{bonden} are consistent with binding energy universality~\cite{Ferrante:1983,Rose:1983} and Pauling's relation between bond-length and energy~\cite{Pauling:Book1960}, and also with ab-initio calculations of light elements.

The angular function is typically given by
\begin{equation}
  g(\theta)
  =
  \gamma
  \left(
  1 + \frac{c^2}{d^2} - \frac{c^2}{d^2 + [h + \cos\theta]^2}
  \right),
  \label{angular1}
\end{equation}
where $c$, $d$ and $h$ are free parameters.
The parameter $K$ characterizes the relationship between equilibrium bond energy and bond length for different crystal structures.

The range of all distance-dependent functions is limited to nearest-neighbors by a switching function $S_{ij}=f_C(r_{ij})$ that depends on the distance between atoms $i$ and $j$ only.
Generally, $f_C$ drops from a value of one to zero between two distances $r_1$ and $r_2$, respectively.
A common choice of cut-off function that is also employed in Refs.~\onlinecite{Tersoff:1989p5566} and \onlinecite{Erhart:2005p035211} uses trigonometric functions and is given by
\begin{equation}
  f_C(r) =
  \left\{
  \begin{array}{ll}
    1 & \text{if } r \leq r_1 \\
    \frac{1}{2} \left[ 1 + \cos\left(\pi\frac{r-r_1}{r_2-r_1}\right) \right]  & \text{if } r_1 < r < r_2 \\
    0 & \text{if } r \geq r_2
  \end{array}
  \right.
  \label{cutoff}
\end{equation}

The interatomic potential that is defined by Eqs.~(\ref{bonden}) to
(\ref{cutoff}) with minor differences in the choices of $\phi(r)$ and
$\beta(r)$ and $b_{ij}$ has been used to parameterize, among others, the
interaction of 
B-C-N,\cite{Albe:1998p111,Matsunga:2000p48}
C-H,\cite{Brenner:1990p9458,Brenner:1992p1948,Brenner:2002p783}
C-O,\cite{Ni:2004p7261}
C-O-H,\cite{Knippenberg:2012p164701}
Ga-As,\cite{Albe:2002p035205}
Fe-C,\cite{Muller:2007p326220,Henriksson:2009p144106}
Pt-C,\cite{Albe:2002p195124}
Si-C,\cite{
Tersoff:1986p632,
Tersoff:1988p6991,
Tersoff:1989p5566,
Erhart:2005p035211}
Si-C-H,\cite{Schall:2012}
Si-O,\cite{Yu:2007p085311,Shan:2010p235302}
W-C-H \cite{Juslin:2005p123520}
and 
Zn-O.\cite{Erhart:2006p6585}
While it is possible to go beyond second moments to higher chemical accuracy, such potentials have only been developed for few element combinations, such as Mo,~\cite{Mrovec:2004p094115} W,~\cite{Mrovec:2007p104119} Fe,~\cite{Mrovec:2011p246402} C-H~\cite{Oleinik:1999p8500,Mrovec:2007p230} and Si.~\cite{Gillespie:2007p155207}

\section{Smoothness of interatomic potentials}

Experience tells that potential energy surfaces as obtained for example from density functional theory (DFT) calculations are smooth.
This fact is underlined by the recent success of using Gaussian processes~\cite{Rasmussen:Book2005} to extrapolate from a finite set of energies obtained from DFT calculations to arbitrary configurations.~\cite{Bartok:2010p136403}
A couple of DFT calculations typically suffice to reconstruct high-accuracy potential energy surfaces.
In Gaussian processes, smoothness is intrinsically programmed into the extrapolation by the covariance function.

\begin{figure}
  \begin{center}
  \includegraphics[width=7cm]{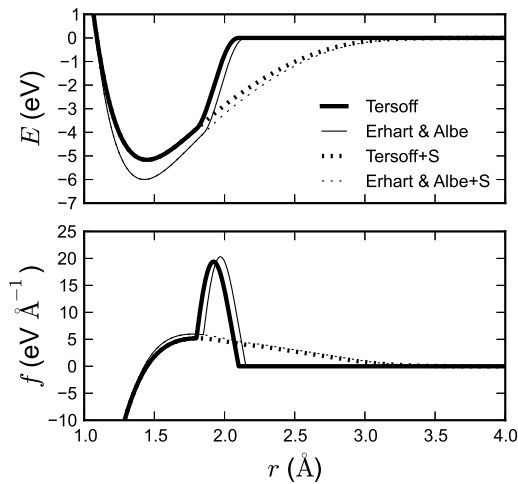}
  \caption{\label{dimer}
    Energy $E$ and tension $f=dE/dr$ as a function of distance $r$ for the carbon dimer.
    Shown are results obtained using the original formulation of the potentials and the screened version (denoted by +S) that unlock the asymptotic behavior of the interaction.
  }
  \end{center}
\end{figure}

The potential energy landscape obtained from Eqs.~(\ref{bonden})-(\ref{cutoff}) is not smooth because the cut-off function Eq.~\eqref{cutoff} forces energies to zero within a short distance interval.
This leads to a failure in the description of transition states that is most easily demonstrated for the dimer.
Fig.~\ref{dimer} shows the energy and tensile force of the carbon dimer as computed using Tersoff's and Erhart \& Albe's potential.
The energy drops to zero steeply as the cut-off is approached.
This leads to an overestimation of the force required to break this bond, with implications for the simulation of cracks and tribology.
We here generalize the meaning of the switching function $S_{ij}$ with $S_{ij}=1$ meaning that a bond exists.
This allows to unlock the asymptotic behavior that is programmed into $\phi(r)$ and $\beta(r)$ for any structure.

As already mentioned above, the cut-off function is designed to allow interaction of nearest neighbors only.
Physically, this can be motivated by the fact that the bond-integral (here the attractive part $\beta$ of the potential) follows a different functional form for second and farther neighbors that is smaller in magnitude.~\cite{NguyenManh:2007p255}
The interaction of second and farther neighbors is ``screened'' by the nearest-neighbor atom.
A central approximation in empirical bond-order potentials is to assume perfect screening for second and farther neighbors and set their bond integral to zero.
This approximation works best for half-filled bands.~\cite{Abell:1985p6184}
In a tight-binding (molecular orbital) picture, the physics of screening functions can be traced back to non-orthogonality.~\cite{Nguyen-Manh:2000p4136}

\subsection{Cut-off procedure}

Besides finding nearest neighbors, a cut-off criterion needs to be able to smoothly interpolate upon transitions that involve changes in coordination number.
We have recently proposed to determine nearest-neighbor relationship~\cite{Pastewka:2008p161402} from the screening function introduced by Baskes et al.~\cite{Baskes:1994p505} that fulfills this condition.
Later, Kumagai et al.~\cite{Kumagai:2009p064310} have proposed an almost identical scheme.

The procedure is a follows: Instead of counting atoms within a certain distance towards a bond, we look for third atoms in the vicinity of the bond.
If any third atom sits close to the bond it is screened, if it sits far away, the bond is allowed to persist.
In this picture a bond is unscreened if there is a line of sight between the two atoms participating in the bond.
A simple empirical and quantitative measure for this intuitive picture is given by constructing ellipsoids of revolution through two atoms.
If a third atom sits inside these ellipsoid the bond is screened.

\begin{figure}
  \begin{center}
  \includegraphics[width=9cm]{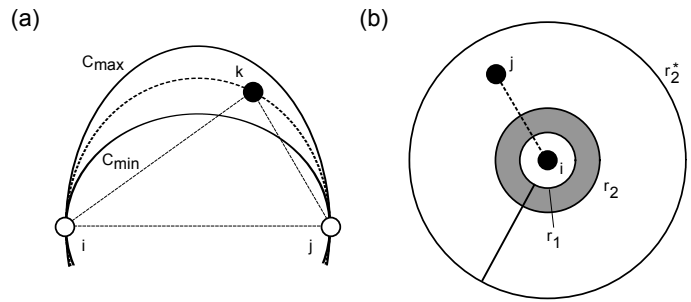}
  \caption{\label{screening}
    (a) Screening of the bond $i$-$j$:
    If an atom $k$ moves into the vicinity of the bond $i$-$j$ we construct an ellipsis through atoms $i$, $j$ and $k$ such that bond $i$-$j$ constitutes one of the half-axes.
    The square root of the coefficient $C_{ijk}$ is then the ratio of the lengths of second half axis to first half axis.
    The $C_{ijk}$ is an empirical measure for how close atom $k$ sits to bond $i$-$j$.
    (b) Our screening approach distinguishes two cutoff radii.
    The bond $i$-$j$ always exists if $r_{ij} < r_1$.
    For $r_2<r_{ij}$ we compute the screening function of panel (a) and determine whether bond $i$-$j$ exists from the value of $C_{ijk}$.
    If an atom sits in the inner gray region where $r_1<r_{ij}<r_2$ we interpolate between the screened and unscreened cutoff function (see Eq.~\eqref{Sinterp}).
    For reasons of computational efficiency we furthermore turn the interaction completely off for $r_{ij}>r_2^*$.
  }
  \end{center}
\end{figure}

Let $r_{ij}$ denote the distance between atom $i$ and atom $j$ for which we would like to compute whether interaction is possible.
We construct an ellipsis through a third atom $k$ (see Fig.~\ref{screening}a). With $X_{ik}=(r_{ik}/r_{ij})^2$ the coefficient
\begin{equation}
  C_{ijk}= \frac{2 (X_{ik}+X_{jk}) - (X_{ik}-X_{jk})^2-1}{1-(X_{ik}-X_{jk})^2}
  \label{Cijk}
\end{equation}
gives the square of the ratio of the two half axes' lengths.
We now consider a bond between atoms $i$ and $j$ to be entirely screened by atom $k$ if the coefficient falls below a critical value $C_\textrm{min}$, while an unscreened bond corresponds to $C_{ijk}>C_\textrm{max}$.
A geometric explanation for the $C_{ijk}$ coefficient is given in Fig.~\ref{screening}.

We now impose the cutoff on the value of $C_{ijk}$ rather than $r_{ij}$.
We define the screening function $\Sigma_{ij}$ of bond $i$-$j$ to be given by $\Sigma_{ij}=0$ if the bond $i$-$j$ is entirely screened and otherwise by~\cite{Baskes:1994p505}
\begin{equation}
  \Sigma_{ij}=\prod_{k, C_{ijk} < C_\textrm{max}}\exp \left[ - \left( \frac{C_\textrm{max}-C_{ijk}}{C_{ijk}-C_\textrm{min}} \right)^2 \right].
  \label{Sij}
\end{equation}
The product runs over all atoms $k$ which are neighbors to the bond $i$-$j$.
For each neighbor $k$ we test whether atom $k$ might screen the bond, and multiply the contributions to the screening function accordingly.
Additionally, we do not want the screening to be active in high pressure situations, where solids may be compressed to highly coordinated structures.
Hence, we define an inner core region where screening is inactive by choosing the switching function to be (see Fig.~\ref{screening}b)
\begin{equation}
  S_{ij} = f_S(r_{ij}) + (1-f_S(r_{ij})) \Sigma_{ij}
  \label{Sinterp}
\end{equation}
Here $f_S$ is a function that drops from unity to zero between radii $r_1$ and $r_2$ where we switch from a bond that cannot be screened to a bond that can be screened by its neighbors.
Note that $\Sigma_{ij}$ is differentiable more than twice.
To make the overall potential energy landscape differentiable more than twice we use:
\begin{equation}
  f_S(r) =
  \left\{
  \begin{array}{ll}
    1 & \text{if } r \leq r_1 \\
    \exp\left[-\left(2\frac{r-r_1}{r_2-r_1}\right)^3\right]
    & \text{if } r_1 < r < r_2 \\
    0 & \text{if } r \geq r_2
  \end{array}
  \right.
  \label{newcutoff}
\end{equation}
This switching procedure does not introduce an additional (artificial) length scale and is intrinsically infinitely ranged.
The ``infinite range'' is manifested by the fact that all distances occurring in Eq.~(\ref{Cijk}) are normalized by the bond distance $r_{ij}$.

\subsection{Long-ranged limits of the bond-order term}

The long-rangedness necessitates an additional modification to traditional empirical bond-order potentials.
The switching function $S$ appears in the total energy Eq.~\eqref{bonden}, but also in the definition of the bond-order Eq.~(\ref{chiij}).
Since for most potentials $\mu=0$ we find $h(r_{ij},r_{ik})=1$ and hence the bond-order $b_{ij}$ becomes independent of the actual bond length and approaches the wrong limit in some situations.
One of these situations occurs when a crystal is cleaved to expose two surfaces.
As we pull the crystal apart to introduce two free surfaces the total energy of the system needs to asymptotically approach the energy of two separated systems.

\begin{figure}
  \begin{center}
  \includegraphics[width=8cm]{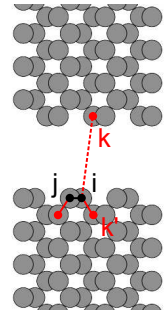}
  \caption{\label{bo_across_surf}(Color online) The bond-order $b_{ij}$ of bond $i$-$j$ that sits at the surface of an exemplary diamond $(110)$ surface depends on all neighbors.
    This includes neighbor $k$ that sits on an opposite surface.
    At sufficient separation the influence of $k$ on $b_{ij}$ needs to vanish.
  }
  \end{center}
\end{figure}

For the specific bond $i$-$k$ shown in Fig.~\ref{bo_across_surf} the bond-length $r_{ik}$ increases continuously with increasing separation $x$.
The values of $\phi(r_{ik})$ and $\beta(r_{ik})$ then drop to zero as $r_{ik}\to\infty$.
However, bond $i$-$j$ feels the presence of atom $k$ in the three body term $b_{ij}$.
For $h=1$, this term is given by
\begin{equation}
  b_{ij} = \left[ 1 + \left( \sum_{\kappa\not=i,j} S_{i\kappa} g(\theta_{ij\kappa}) \right)^\eta \right]^{-\delta},
  \label{boexample}
\end{equation}
and independent of the absolute length $r_{i\kappa}$ of bond $i$-$\kappa$ if that bond is unscreened and $S_{i\kappa}=1$.
For the particular bond $i$-$k$ shown in Fig.~\ref{bo_across_surf} we have $S_{ik}=1$.
Without any mechanism to eliminate the influence of atom $\kappa=k$ to the bond-order in $b_{ij}$ in Eq.~\eqref{boexample} the bottom surface will feel the top surface's presence at arbitrary distances since $S_{ik}=1$.
Without screening functions we have $S_{ik}=f_C(r_{ik})$ depend only on distance, and the contribution of $k$ will have vanished once the atom has moved out of the cut-off radius of atom $i$, i.e. once $r_{ik}>r_2$.

Here we argue that in order to provide a well defined limiting value for the bond-order with increasing bond-length we must choose $\mu>0$.
The exponential term $h(r_{ij},r_{ik})$ then provides the necessary asymptotics of the bond-order at large distances.
In the above example, the contribution of bond $i$-$k$ to $b_{ij}$ will decay exponentially as $r_{ik}$ increases.
Usually, $\mu$ is treated as an adjustable parameter, but tight-binding bond theory tells us that for an expansion up to second moments and ignoring the contribution of $\pi$-orbitals the total energy needs to be~\cite{Horsfield:1996p12694}
\begin{equation}
  E = \frac{1}{2}\sum\limits_{i<j} S_{ij} \left[
  \phi(r_{ij}) - \frac{\beta^2(r_{ij})}{\sqrt{\beta^2(r_{ij})+\sum\limits_{k\not=i,j}S_{ik} \beta^2(r_{ik}) g(\theta_{ijk})}}
  \right].
  \label{bonden2}
\end{equation}
This is compatible with the empirical Tersoff-Brenner formulation if we choose $\eta=1$, $\delta=1/2$ (see also Ref.~\onlinecite{Abell:1985p6184}) and
\begin{equation}
  h(r_{ij}, r_{ik})
  =
  \left(\frac{\beta(r_{ik})}{\beta(r_{ij})}\right)^2.
  \label{hbeta}
\end{equation}
Using the functional form Eq.~\eqref{h} for $h(r_{ij},r_{ik})$ and Eq.~(\ref{beta}) for $\beta(r)$ we obtain $m=1$ and
\begin{equation}
  \mu = \alpha \sqrt\frac{2}{K}.
  \label{mu}
\end{equation}
%

Unfortunately, for $m=1$ the value of $\mu$ contributes to the $C_{44}$ shear modulus of the material.
This is easily seen from the definition of this particular modulus:
$C_{44}$ is given by~\cite{Finnis:Book2004}
\begin{equation}
  C_{44}=\frac{1}{3V} \frac{\partial^2 E}{\partial\epsilon^2}
  \label{C44}
\end{equation}
where $V$ is the volume of the crystal and $E$ its total energy.
The strain $\epsilon$ characterizes the shear transformation where all atoms are transformed from position $\v{r}$ to $\v{r}' = (1+\t{T}) \v{r}$ with
\begin{equation}
  \t{T} = \left( \begin{array}{ccc}
    0 & \epsilon/2 & \epsilon/2 \\
    \epsilon/2 & 0 & \epsilon/2 \\
    \epsilon/2 & \epsilon/2 & 0
    \end{array} \right)
\end{equation}
This particular transformation stretches some bonds in the diamond structure and contracts others.
The second derivative of Eq.~\eqref{C44} then involves terms such as:
\begin{equation}
  \begin{split}
  \frac{\partial^2 h}{\partial r_{ij} \partial r_{ik}}
  =&
  -m(m-1) (2\mu_{ik})^m(r_{ij}-r_{ik})^{m-2} h(r_{ij},r_{ik})
  \\
  &-m^2 (2\mu_{ik})^{2m} (r_{ij}-r_{ik})^{2m-2} h(r_{ij},r_{ik})
  \end{split}
\end{equation}
In the equilibrium diamond structure $r_{ij}=r_{ik}$ and this derivative vanishes only if $m>2$.
Choosing $m=1$ would hence require a complete readjustment of all parameters to a set of material properties.
For small deviations from the crystalline ground-state, $m=3$ hence removes the contribution of $\mu$ to the energy.
While this is not fully consistent with Eq.~\eqref{hbeta}, we use $m=3$ in the following for convenience and to avoid refitting the potential.
Since $\mu$ needs to have units of inverse length, we empirically choose $\mu_{ik}=r_0^{-1}$ to be the inverse of the dimer length $r_0$ of elements $i$-$k$.

The Silicon potential of Kumagai et al. has a value of $\mu\not=0$ that is independently fit.
Here we therefor retain $m=1$.
Note that Kumagai et al. fit $\mu=1.8\,\text{\AA}^{-1}$ while from Eq.~\eqref{mu} we obtain a value of $\mu=1.4\,\text{\AA}^{-1}$.
We also note here that in our earlier screened REBO2 potential we enforced the proper limiting behavior for $b_{ij}$ by an additional cutoff function $h(r_{ij}, r_{ik}) = f_C^h(r_{ik})$ that depended on the distance $r_{ik}$ only.
For the potentials presented in this paper, we use the expression given by Eq.~\eqref{h} because we believe that choosing a functional form close to that given by tight-binding bond theory is crucial for the transferability of the interatomic potential.

\subsection{Computational considerations}

A full cut-off free formulation as presented in the preceding chapters is possible by computing a Voronoi tessellation of the atomic configuration in each time step.
The screening functions would then be computed for atoms whose respective Voronoi cells share a face.
However, this approach is computationally expensive and not linear scaling.
In all practical cases, we therefor smoothly cut the interaction off at a certain distance to be able to use the usual linear scaling linked cell algorithms.~\cite{Allen:Book1989}
If this distance is large, the modulation of the bond-integrals will be weak and their asymptotic behavior essentially conserved.
The final expression for the switching function we use is hence
\begin{equation}
  S_{ij} = f_S(r_{ij}) + (1-f_S(r_{ij})) f_C^*(r_{ij}) \Sigma_{ij}
  \label{Sinterp2}
\end{equation}
with $f_C^*(r)=f_S(r)$ that switches between radii $r_1^*$ and $r_2^*$.

The specific parameters for the potentials presented in this article are given in Tab.~\ref{params}.
The parameters are chosen with the following considerations in mind: $r_1$ and $r_2$ must lie between the first and second neighbor shell in the diamond or 3C structure (for C, Si and Si-C) and between the first and second neighbor shell in graphite (for C).
Furthermore, $r_1$ and $r_2$ for Si-Si must be smaller than the first Si-Si neighbor shell in 3C Si-C.
The latter constraint is the reason why $r_1$ and $r_2$ for Si-Si are smaller than for C-C and Si-C if compared to the crystalline bulk bond length.
The outermost cutoff $r_2^*$ must be large enough to eliminate spurious peaks in the dimer force curves and the cohesive stress functions discussed below.
This is usually achieved at about $r_2^*\approx 2.5 r_\text{nn}$ where $r_\text{nn}$ is the nearest neighbor distance in the diamond or 3C structure.
We furthermore empirically fix $r_2=1.2 r_1$ and $r_2^*=2 r_1^*$.
For the Tersoff potential we use the original Tersoff-Lorentz-Berthelot~\cite{Tersoff:1989p5566} mixing rule $r_\text{SiC}=\sqrt{r_\text{C} r_\text{Si}}$ for $r_1$ and $r_1^*$.
The values of $C_\text{min}$ and $C_\text{max}$ are chosen such that for three atoms located on the corners of an equilateral triangle three unscreened bonds exist, and for four atoms on the corners of a square four bonds exist.

\begin{table}
\begin{tabular}{l@{\extracolsep{7mm}}ccc@{\extracolsep{2mm}}c}
  \hline\hline
    & C-C & Si-Si & \multicolumn{2}{c}{Si-C} \\
  \hline\hline \\
  & \multicolumn{2}{c}{all potentials} & TIII+S & EA+S \\
  \hline
  $r_1$ (\AA) & $2.00$ & $2.50$ & $2.24$ & $2.40$ \\
  $r_2=1.2r_1$ (\AA) & $2.40$ & $3.00$ & $2.68$ & $2.88$ \\
  $r_1^*$ (\AA) & $2.00$ & $3.00$ & $2.45$ & $2.40$ \\
  $r_2^*=2r_1^*$ (\AA) & $4.00$ & $6.00$ & $4.90$ & $4.80$
  \vspace{1mm} \\
  $C_\text{min}$ & \multicolumn{4}{c}{--- $1.0$ ---} \\
  $C_\text{max}$ & \multicolumn{4}{c}{--- $3.0$ ---}
  \vspace{5mm} \\
  & \multicolumn{4}{c}{TIII+S} \\
  \hline
  $\mu=r_0^{-1}$ (\AA$^{-1}$) & $0.69$ & $0.57$ & \multicolumn{2}{c}{$0.44$}
  \vspace{5mm} \\
  & \multicolumn{4}{c}{EA+S} \\
  \hline
  $\mu=r_0^{-1}$ (\AA$^{-1}$) & $0.70$ & $0.56$ & \multicolumn{2}{c}{$0.54$}
\end{tabular}
\caption{\label{params}Parameters for the screened Tersoff (TIII), Erhart \& Albe (EA) and Kumagai potentials.}
\end{table}

\section{Properties of the screened potentials}

We report some select properties of the screened potentials and compare those to their unscreened counterparts and higher level quantum calculations.~\cite{code}
In what follows, we denote Tersoff's third-generation potential~\cite{Tersoff:1989p5566} as TIII, and the screened incarnation as TIII+S.
Similarly, we denote Erhart \& Albe's potential~\cite{Erhart:2005p035211} as EA, and the screened incarnation as EA+S.
Kumagai et al.'s~\cite{Kumagai:2007p457} potential will be referred to as Kumagai and Kumagai+S in it's unscreened and screened incarnation, respectively.
For completeness, we also compare to results obtained with the REBO2~\cite{Brenner:2002p783} and screened REBO2 (REBO2+S)~\cite{Pastewka:2008p161402} potential for carbon, and the Stillinger-Weber (SW) potential for silicon.
%

If not otherwise noted, DFT reference calculations are carried out by us and employ the local density approximation~\cite{Perdew:1992p13244} and projector augmented waves.~\cite{Blochl:1994p17953}
The wave functions are expanded on a real space grid.
We use the \text{GPAW} code.~\cite{Mortensen:2005p035109,Enkovaara:2010p253202}
Table~\ref{SiC_props} lists some properties of diamond, silicon and 3C silicon carbide as obtained from the classical potentials and this particular DFT method.

\begin{table*}
\begin{tabular}{l@{\extracolsep{7mm}}c@{\extracolsep{7mm}}c@{\extracolsep{3mm}}c@{\extracolsep{7mm}}c@{\extracolsep{3mm}}cc@{\extracolsep{7mm}}c}
\multicolumn{7}{c}{diamond}
\\\hline\hline
& Expt. & \multicolumn{2}{c}{DFT-LDA} & \multicolumn{2}{c}{BOP}
\\
& & & this work & TIII+S & EA+S & REBO2+S
\\\hline
$E_c$ (eV)                   & $-7.37^a$ & $-9.03^e$ & $-8.95$ & $-7.371$ & $-7.373$ & $-7.370$ \\
$a_0$ (\AA)                  & $3.567^b$ & $3.528^e$ & $3.535$ & $3.566$ & $3.566$ &  $3.566$
\vspace{1mm}\\
$C_{11}$ (GPa)               & $1076^c$ & $1060^f$ & $1094$   & $1074$    & $1088$  &  $1076$ \\
$C_{12}$ (GPa)               & $125^c$ & $125^f$ & $147$    & $102$     & $125$   &  $125$ \\
$C_{44}$ (GPa)               & $577^c$ & $562^f$ & $584$    & $641$     & $641$   &  $720$ \\
$C_{44}^0$ (GPa)             & & & $591$    & $671$    & $673$   & $738$
\vspace{1mm}\\
$\gamma_{\{111\}}$ (J m$^{-2}$)           & $5.3^d$ & $6.43^g$ & $6.37$ & $2.75$ & $2.06$ & $5.37$ \\
$\gamma_{\{110\}}$ (J m$^{-2}$)           & $6.5^d$ & $5.93^g$ & $5.90$ & $4.04$ & $2.96$ & $3.12$ ($3.35$) \\
$\gamma_{\{100\}}$ (J m$^{-2}$)           & $9.2^d$ & $9.40^g$ & $9.34$ & $7.09$ ($6.66$) & $5.88$ ($5.59$) & $7.84$ ($11.0$) \\
$\gamma^{2\times 1}_{\{100\}}$ (J m$^{-2}$) & $-$ & $5.71^g$ & $5.43$ & $6.61$ ($6.33$) & $5.93$ ($5.65$) & $5.27$ ($6.16$)
\vspace{5mm}
\\
\multicolumn{7}{c}{graphite}
\\\hline\hline
& Expt. & \multicolumn{2}{c}{DFT-LDA} & \multicolumn{2}{c}{BOP}
\\
& & & this work & TIII+S & EA+S & REBO2+S
\\\hline
$E_c$ (eV)   & $-7.374^m$ & $-8.61^e$ & $-8.93$ & $-7.395$ ($-7.396$) & $-7.374$ & $-7.414$ ($-7.395$) \\
$a_0$ (\AA)  & $2.461^h$ & $2.440^e$ & $2.445$ & $2.530$ & $2.555$ &  $2.458$ ($2.460$)
\\
$c_0$ (\AA)  & $6.710^h$ & $6.681^e$ & $6.532$ & $[6.710]^p$ & $[6.710]^p$ & $[6.710]^p$
\vspace{5mm}
\\
\multicolumn{8}{c}{silicon}
\\\hline\hline
& Expt. & \multicolumn{2}{c}{DFT-LDA} & \multicolumn{3}{c}{BOP} & SW
\\
& & & this work & TIII+S & EA+S & Kumagai+S
\\\hline
$E_c$ (eV)                 & $-4.62^i$ & $-4.63^c$ & $-4.75$ & $-4.630$ & $-4.628$ & $-4.630$ \\
$a_0$ (\AA)                & $5.431^b$ & $5.400^c$ & $5.406$ & $5.432$ & $5.429$ & $5.429$
\vspace{1mm}\\
$C_{11}$ (GPa)              & $166^b$ & $159^k$ & $160$   & $143$   & $169$    & $166$ \\
$C_{12}$ (GPa)             & $64^b$ & $61^k$ & $63$    & $75$    & $64$     & $65$ \\
$C_{44}$ (GPa)              & $80^b$ & $85^k$ & $82$    & $69$    & $60$     & $77$ \\
$C_{44}^0$ (GPa)            & & $111^k$ & $112$   & $119$   & $105$    & $121$
\vspace{1mm}\\
$\gamma_{\{111\}}$ (J m$^{-2}$)           & $1.23^j$ & $1.74^g$ & $1.72$   & $1.20$ & $1.00$ & $0.89$ & $1.36$ \\
$\gamma_{\{110\}}$ (J m$^{-2}$)$^o$           & $1.510^j$ & $1.70^g$ & $1.68$   & $1.52$ & $1.23$ & $1.08$ & $1.67$ \\
$\gamma_{\{100\}}$ (J m$^{-2}$)            & $2.130^j$ & $2.39^g$ & $2.37$  & $2.16$ ($2.27$)  & $1.90$ ($1.95$)  & $1.70$ ($1.77$) & $2.35$ \\
$\gamma^{2\times 1}_{\{100\}}$ (J m$^{-2}$) & & $1.45^g$ &  $1.53$  & $1.48$ & $1.13$ & $1.07$ & $1.44$
\vspace{5mm}
\\
\multicolumn{7}{c}{3C silicon-carbide}
\\\hline\hline
& Expt. & \multicolumn{2}{c}{DFT-LDA} & \multicolumn{2}{c}{BOP}
\\
& & & this work & TIII+S & EA+S &
\\\hline
$E_c$ (eV)                  & $-6.34^l$ & $-7.42^n$ & $-7.37$ & $-6.165$ & $-6.339$ & \\
$a_0$ (\AA)                 & $4.358^m$ & $4.344^n$ & $4.338$ & $4.321$ & $4.359$ & 
\vspace{1mm}\\
$C_{11}$ (GPa)              & $390^l$ & $390^n$ & $405$    & $437$   & $383$  &  \\
$C_{12}$ (GPa)              & $142^l$ & $134^n$ & $145$   & $118$   & $144$  &  \\
$C_{44}$ (GPa)              & $256^l$ & $253^n$ & $247$    & $257$   & $240$  &  \\
$C_{44}^0$ (GPa)            & & $273^n$ & $279$    & $311$   & $305$   & 
\vspace{1mm}\\
$\gamma_{\{111\}}$ (J m$^{-2}$)           & & & $4.17^q$ & $1.85^q$ & $1.67^q$ & \\
$\gamma_{\{110\}}$ (J m$^{-2}$)           & & & $3.29$ & $2.40$ & $2.29$ & \\
$\gamma_{\{100\}}$ (J m$^{-2}$)           & & & $5.46^q$ & $4.12$ ($4.21$)$^q$ & $3.87$ ($3.93$)$^q$ & \\
$\gamma^{2\times 1}_{\{100\}}$ (J m$^{-2}$) & & & $3.48^q$ & $2.87$ ($2.85$)$^q$ & $2.92$ ($2.85$)$^q$ & 
\end{tabular}
\caption{\label{SiC_props}Properties of diamond, silicon and 3C silicon-carbide. Unless referenced, DFT results are LDA (see text). Values in parenthesis are for the unscreened potential, if different from their screened counterpart.
$C_{44}^0$ is the $C_{44}$ modulus obtained without relaxation of atomic positions.
The $(111)$ surface is cut at the shuffle plane.
$^a$Ref.~\onlinecite{Yin:1981p6121}
$^b$Ref.~\onlinecite{CRC93}
$^c$Ref.~\onlinecite{Grimsditch:1975p3139}
$^d$Fracture energy, Ref.~\onlinecite{Ramaseshan:1946p114}
$^e$Ref.~\onlinecite{Furthmuller:1994p15606}
$^f$DFT-GGA, Ref.~\onlinecite{Mounet:2005p205214}
$^g$Ref.~\onlinecite{Stekolnikov:2002p115318}
$^h$Refs.~\onlinecite{Boettger:1997p11202,Zhao:1989p993,Hanfland:1989p12598}
$^i$Ref.~\onlinecite{Farid:1991p14248}
$^j$Ref.~\onlinecite{Jaccodine:1963p524}
$^k$Ref.~\onlinecite{Nielsen:1985p3792}
$^l$Ref.~\onlinecite{Lambrecht:1991p3685}
$^m$Ref.~\onlinecite{Landolt-Bornstein}
$^n$Ref.~\onlinecite{Karch:1994p17054}
$^o$The Si $(110)$ surface reconstructs in DFT-LDA. The empirical potentials do not capture this reconstruction.
$^p$Fixed to the experimental value.
$^q$Energies obtained by creating a silicon and a carbon terminated surface.}
\end{table*}

\subsection{Fracture}

\subsubsection{Cohesive stress}

\begin{figure*}
  \begin{center}
  \includegraphics[width=18cm]{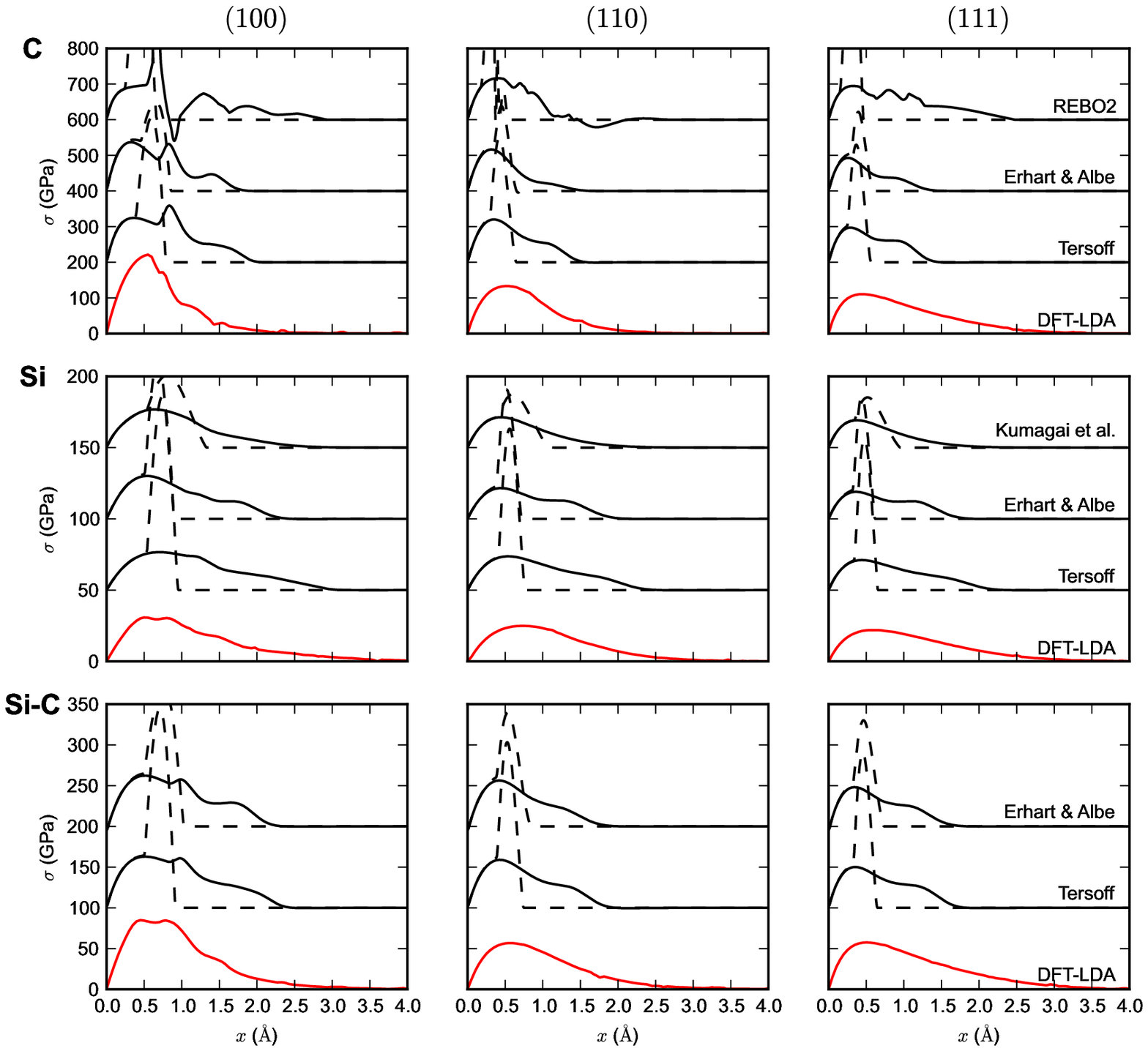}
  \caption{\label{cohesive_forces}(Color online)
  Cohesive stress $\sigma$ (derivative of the energies obtained by separating a bulk crystal normal to certain surfaces normalized by their area) for carbon, silicon and 3C silicon-carbide.
  Here we show this function for the creation of low-index $(100)$, $(110)$ and $(111)$ surfaces.
  Broken lines show results for the unscreened interatomic potentials.
  Solid lines show their screened counterparts and DFT results.
  Curves are shifted vertically to be distinguishable.
  The $(111)$ surface is cut at the shuffle plane.
  }
  \end{center}
\end{figure*}

We compute the cohesive stress functions by separating the ideal bulk of the crystal for diamond, silicon, and 3C silicon-carbide to create $(100)$, $(110)$ and $(111)$ surfaces.
These calculations are carried out unrelaxed, and the cohesive stress function that is shown in Fig.~\ref{cohesive_forces} is the first derivative of the total energy curves obtained normalized by the exposed surface area.
The value of $x$ denotes the distance of the newly created surface such that $x=0$ is the limit of the bulk crystal and $x\to\infty$ are two free surfaces.

The maximum force obtained for all structures and all surfaces probed here is in reasonable agreement with the DFT calculations.
However, the asymptotic behavior of the cohesive stress is significantly lower than the values obtained from DFT calculations for the $(110)$ and even worse for the $(111)$ surface.
Since the surface energy is the area beneath the cohesive stress functions of Fig.~\ref{cohesive_forces}, this difference can be attributed solely to a mismatch in surface energy.
We list the energies for these surfaces in Tab.~\ref{SiC_props}.
While all potentials give reasonable values for the high energy $(100)$ surface, the agreement with DFT calculation for the $(110)$ and $(111)$ surfaces are worse.
For carbon and 3C silicon-carbide, the order of the energetics of $(110)$ and $(111)$ surfaces is reversed in DFT calculations and experiments.
All potential except for the REBO2+S follow the experimental order.
It is somewhat surprising that these classical potentials appear to capture the peak force at the transition state more accurately than the equilibrium surface energies.

For opening a $(100)$ diamond surface we find that the force for TIII+S and EA+S has two distinct peaks, with the peak at the larger separation having a higher force.
These peaks are less pronounced, but also visible, for the silicon-carbide $(100)$ surface, but do not show up on the $(110)$ or $(111)$ surfaces.
The origin of this is a too sudden drop in $\chi_{ij}$ that stems from choosing $m=3$, and not $m=1$ in Eq.~\eqref{chiij} as the rigorous bond-order theory suggests.~\cite{Horsfield:1996p12694}
This in return leads to an overestimation of $b_{ij}$ for the transition state and hence a potential that is too attractive in that region.

Finally, we note that the potential energy landscape of the REBO2+S is more corrugated than the one obtained for TIII+S, EA+S and Kumagai+S.
This is related to the treatment of $\pi$-electron in the REBO formalism.
In brief, an additive correction is applied to $b_{ij}$ and $\chi_{ij}$ (given in Eqs.~\eqref{bij} and \eqref{chiij}, respectively).
The value of that correction depends on the coordination numbers of the atoms in the vicinity of the bond and was fit to the atomization energies of a select set of hydrocarbon molecules.
Since coordination numbers are integer values, the transition values upon changes in coordination are obtained from a cubic spline interpolation.
This cubic spline is the origin of the additional corrugation seen for the REBO2+S in Fig.~\ref{cohesive_forces}.
We also note that for $(110)$ and $(111)$ surfaces the coordination number jumps from $4$ for a bulk atom to $3$ for a surface atom.
On the $(100)$ surface, the coordination number jumps from $4$ in the bulk to $2$ at the surface giving rise to an additional transition state with coordination number $3$ that is the origin of the peaks seen in Fig.~\ref{cohesive_forces} for REBO2+S on this particular surface.
The simpler formulation given by Eqs.~(\ref{bonden}) to (\ref{cutoff}) without the spline corrections that is the basis of the TIII, EA and Kumagai potentials has the advantage that it yields a smoother potential energy landscape, albeit at the cost of limited accuracy in particular in complex molecular systems.

\subsubsection{Static crack}

In addition to the cohesive stress function we compute bond-breaking events in a mode I crack geometry using the method by P\'erez and Gumbsch.~\cite{Perez:2000p5347,Perez:2000p4517}
In brief, we consider a small atomistic region around the crack tip and fix the boundary atoms of this region using the near field solution of the displacements from linear elastic fracture mechanics.
Then, the stress intensity factor $K$ is increased step-wise, the system is relaxed,~\cite{Bitzek:2006p170201} and we monitor the length of the bond in front of the crack tip.
We also investigate the closing of a crack by decreasing the stress intensity factor and monitoring the length of the bond behind the crack tip.
In all calculations the crack tip is centered on the bond of interest.
More information on the technique can be found in Refs.~\onlinecite{Perez:2000p5347} and \onlinecite{Perez:2000p4517}.

\begin{figure}
  \begin{center}
  \includegraphics[width=7cm]{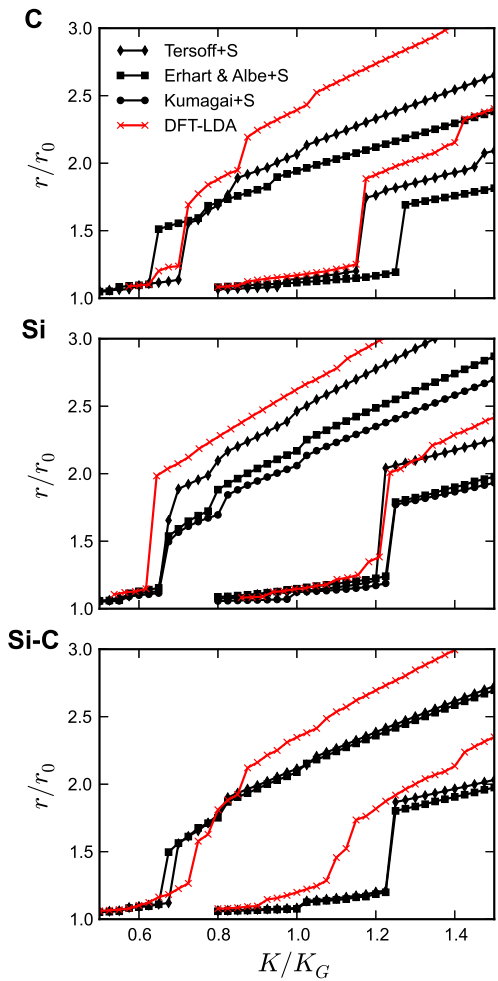}
  \caption{\label{cracks}(Color online)
  Bond length $r$ before the crack tip (left three curves) and after the crack tip (right three curves) for a crack on the $(110)$ surface with a $[1\bar{1}0]$ crack front in carbon, silicon and 3C silicon-carbide as a function of the stress intensity factor $K$.
  The bond length $r$ is here scaled by the bond length $r_0$ in the bulk crystal, and the stress intensity factor is scaled by the factor $K_G$ obtained from Griffith's criterion using relaxed surface energies.
  }
  \end{center}
\end{figure}

Results for a crack on the $(110)$ surface with a $[1\bar{1}0]$ crack front for diamond, silicon and 3C silicon-carbide are shown in Fig.~\ref{cracks}.
We do not show the unscreened potentials which do not break bonds in this kind of simulation.
For TIII+S and EA+S the agreement with DFT calculations is reasonable.
For diamond and silicon, the TIII+S follows the DFT results almost exactly in predicting the correct stress intensity factor $K$ for bond breaking and bond formation.
EA+S overestimates the stress intensity factor $K_+$ required for breaking and underestimates the stress intensity $K_-$ for bond formation hence giving a too large lattice trapping region $\Delta K=K_+-K_-$ for carbon.
For silicon, the width of the lattice trapping region $\Delta K$ is well described by both potentials.
For 3C silicon-carbide, TIII+S and EA+S give almost identical results but overestimate the lattice trapping $\Delta K$.
Additionally, the opening of the bond in our DFT calculations proceeds more smoothly.
This could be related to charge transfer that occurs in silicon-carbide and is not captured by our potentials.

\subsection{Melting}

We determine the melting point for diamond, silicon and 3C silicon-carbide by equilibrating a crystal-melt interface in a simulation without heat exchange with some external bath.
In these simulations, the crystal-melt interface advances or recedes until the system is equilibrated to the melting temperature.
In all cases the $(100)$ surface is exposed to the melt and the pressure is controlled by an anisotropic Andersen barostat~\cite{Andersen:1979p2384} that controls the box size independently in all three Cartesian directions.

The result of this calculation are summarized in Tab.~\ref{melting_points}.
The melting points for diamond are taken at the pressure of the diamond/graphite/melt triple point ($12.4\,\text{GPa}$) and lie in the range of experimental values for all screened potentials.
We note that while $4713\,\text{K}$ seems to be the universally referenced melting point of diamond, the experimental values spread over a much larger range with initial melting reported at temperatures as low as $3500\,\text{K}$.~\cite{Bundy:1996p141}
No melting point could be obtained for the unscreened potential because bulk diamond spontaneously transforms into a graphite under these pressure/temperature conditions.
Since the interaction range of the unscreened case is considerably smaller than the interlayer graphite spacing this conversion can proceed without a volume expansion and hence without performing work against the external pressure.
The screened potential have a longer range.
The individual graphitic sheets do interact and inhibit this transition at sufficiently high pressures.

\begin{table}
\begin{tabular}{lccc}
& C$^a$ & Si & 3C-Si-C \\
\hline
Expt. & $3500$---$5000^{b}$, $4713^{c}$ & $1687^e$ & $2818\pm 40^e$
\vspace{1mm}
\\
TIII & $-^d$ & $2580\pm 15$ \\
TIII+S & $5240\pm 30$ & $2330\pm 15$ & $3190 \pm 15$
\vspace{1mm} \\
EA & $-^d$ & $2510\pm 15$ \\
EA+S & $4210\pm 25$ & $2365\pm 15$ & $3235\pm 15$
\vspace{1mm} \\
REBO2 & $-^d$ & $-$ & $-$ \\
REBO2+S & $3950\pm 20$ & $-$ & $-$
\vspace{1mm} \\
Kumagai & $-$ & $1725\pm 10$ & $-$ \\
Kumagai+S & $-$ & $1625\pm 10$ & $-$
\vspace{1mm} \\
SW & $-$ & $1636\pm 5^f$ & $-$
\end{tabular}
\caption{\label{melting_points}Melting points of crystalline diamond, silicon 3C-Si-C in Kelvin for the different potentials studied here. The melting point was determined by equilibrating a $(100)$ surface with the melt. Error is the standard deviation of the temperature fluctuation in an equilibrated NVE ensemble of $18432$ atoms. $^a$At $12.4\,\text{GPa}$. $^b$Ref.~\onlinecite{Bundy:1996p141} $^c$Ref.~\onlinecite{CRC93} $^d$A diamond/melt interface is unstable in the unscreened potentials, see text. $^e$Ref.~\onlinecite{Olesinski:1984p486}  $^f$System size of $65536$ atoms total.}
\end{table}

The melting point for silicon at zero pressure is overestimated by about $1000\,\text{K}$ by both the TIII and EA potentials.
This overestimation has been noted before,~\cite{Cook:1993p7686,Erhart:2005p035211} and Kumagai and co-workers pointed out that it is related to the angular term.~\cite{Kumagai:2007p457}
The Kumagai potential employs a different angular term and does correct the melting point as independently confirmed by Schelling~\cite{Schelling:2008p274} and here.
However, the improved melting point comes at an expense of surface energies that are considerably lower than TIII and EA energies which themselves are an underestimation of the respective DFT results (see Tab.~\ref{SiC_props}).
For silicon-carbide we obtain melting points that are only about $400\,\text{K}$ too high.
TIII and EA solids melt at roughly identical temperatures.
In all cases, the screening function lowers the melting point compared to the respective unscreened potential by about $10$ to $15\%$.

\subsection{Glass formation}

\subsubsection{Hybridization of amorphous carbon}

Classical empirical bond-order potentials notoriously fail at describing the properties of amorphous carbon that is quenched from the melt.
One particular property that is also accessible from experiments and \emph{ab-initio} calculations it the fraction of diamond-like, sp$^3$ hybridized atoms as a function of the density of the amorphous sample.
For example, the Tersoff, REBO,~\cite{Brenner:1990p9458,Brenner:1992p1948} and REBO2 potentials are known to fail to describe this relationship and typically yield $20\%$ to $40\%$ of sp$^3$ close the the density of diamond where the sp$^3$ fraction should saturate.~\cite{Marks:2000p035401,Marks:2002p2901,Pastewka:2008p161402}
For deposition processes, a common cure is to slightly increase the cut-off range of the potential but keeping it between the first and second nearest neighbor shell of graphite and diamond.~\cite{Jaeger:2000p1129,Jager:2003p24201}
This cure only works above a certain density.~\cite{Mrovec:2007p230}

Here, we compute sp$^3(\rho)$ curves by quenching liquid carbon within $0.5\,\text{ps}$ from $5000\,\text{K}$ to $300\,\text{K}$ at constant volume.
The same procedure has been used in \emph{ab-initio}~\cite{McCulloch:2000p2349} and non-orthogonal tight-binding (NOTB)~\cite{Pastewka:2008p161402} calculations that will be used as a reference here.
We also report the experimental analysis of physically deposited amorphous carbon of Ref.~\onlinecite{Ferrari:2000p11089} for comparison.
An atom contributes towards the sp$^3$ fraction if it has four neighbors within a distance of $1.85\,\text{\AA}$.

\begin{figure}
  \begin{center}
  \includegraphics[width=7cm]{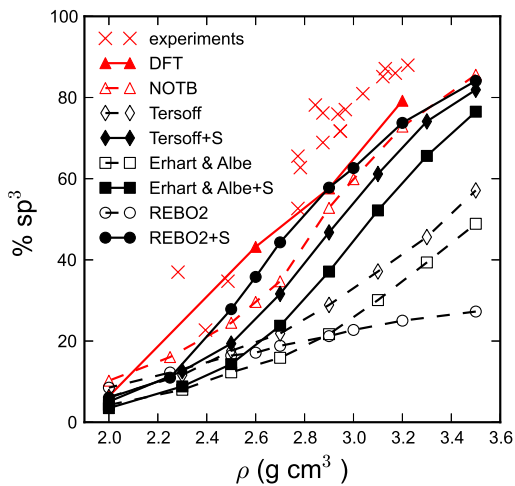}
  \caption{\label{hyb}(Color online)
    Carbon: Fraction of sp$^3$ in amorphous carbon that is quenched from a $5000\,\text{K}$ melt to $300\,\text{K}$ with a time constant of $0.5\,\text{ps}$.
    Experimental data is from Ref.~\onlinecite{Ferrari:2000p11089}, DFT data is from Ref.~\onlinecite{McCulloch:2000p2349} and NOTB data is from Ref.~\onlinecite{Pastewka:2008p161402}.
  }
  \end{center}
\end{figure}

All this data, along with results for the screened and unscreened potentials discussed in this work are shown in Fig.~\ref{hyb}.
The TIII+S potential follows the NOTB data almost exactly, albeit yielding an sp$^3$ fraction that is lower by a few percent.
EA+S also follow the NOTB curve, but the sp$^3$ fraction is lower than the one obtained by TIII+S.
All unscreened potentials are worse, predicting at best $50\%$ to $60\%$ sp$^3$ at densities of $3.5\,\text{g cm}^{-3}$ where the sp$^3$ fraction should be around $80\%$.

\subsubsection{Supercooling amorphous silicon from the melt}

In computer simulations, amorphous silicon is typically quenched from the melt at constant pressure rather than constant volume.~\cite{Demkowicz:2004p025505,Demkowicz:2005p245206,Demkowicz:2005p245205}
We here carry out such simulation at zero external pressure and quench rates of $1\,\text{K}\,\text{ps}^{-1}$ using Berendsen temperature and pressure control~\cite{Berendsen:1984p3864} with relaxation time constants of approximately $1\,\text{ps}$ for temperature and $10\,\text{ps}$ for pressure.
The quench starts from the melt equilibrated at $3000\,\text{K}$.

\begin{figure}
  \begin{center}
  \includegraphics[width=8cm]{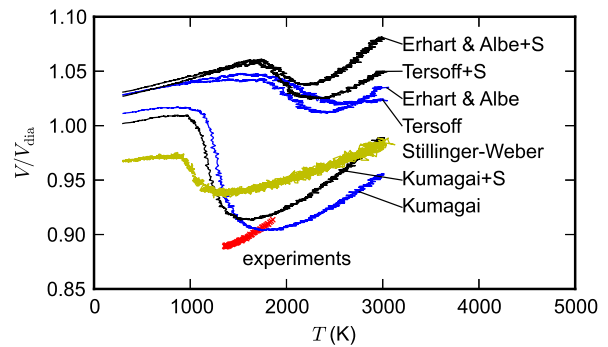}
  \caption{\label{quench}(Color online)
    Silicon: Volume per atom as a function of temperature when quenching from the liquid phase to $300\,\text{K}$ at a rate of $1\,\text{K}\,\text{ps}^{-1}$ and zero external pressure.
    The periodic cell contained $4001$ atoms in all cases.
    Experimental data is taken from Ref.~\onlinecite{Rhim:2000p313}.
  }
  \end{center}
\end{figure}

We first note that the density of the melt does notably depend on the potential under consideration.
Fig.~\ref{quench} shows the atomic volume $V$ as a function of temperature $T$ during the quench.
The volume at the highest temperature ($3000\,\text{K}$) is the equilibrated melt.
All potentials but the screened TIII+S predict a liquid phase that is denser than the supercooled amorphous that is shown at $300\,\text{K}$.
However, only Kumagai, Kumagai+S and the SW potential predict a liquid phase that is denser than the crystalline.
The densest liquid phase is given by the Kumagai potential which is the only potential to reproduce a liquid phase density consistent with experiments.~\cite{Rhim:2000p313}
The screened and unscreened Kumagai potential give similar results.
The screened Kumagai however seems to be marginally better at reproducing the slope $dV/dT$ of the experimental temperature dependence that was reported by Rhim~et~el.~\cite{Rhim:2000p313}

The temperature at which the density peaks during solidification is typically associated with the glass transition temperature $T_g$.~\cite{Demkowicz:2005p245205}
Both Kumagai and SW potentials give a $T_g$ of about $1000\,\text{K}$ in excellent agreement with measurements.~\cite{Hedler:2004p804}
TIII and EA overestimate both glass transition and melting temperature by about a factor of $1.5$.

\subsubsection{Pair distribution functions of amorphous silicon-carbide}

\begin{figure}
  \begin{center}
  \includegraphics[width=8cm]{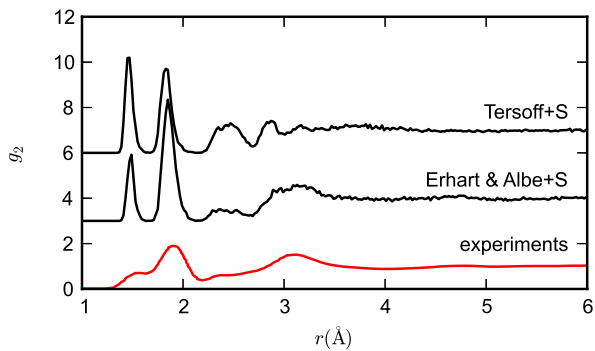}
  \caption{\label{g2}(Color online) Silicon-carbide: Pair distribution functions for stochiometric silicon carbide.
    The final amorphous structure was obtained by quenching a box of $4000$ atoms from the melt to $300\,\text{K}$ at a rate of $1\,\text{K}\,\text{ps}^{-1}$ and zero external pressure.
    Curves are shifted vertically to be distinguishable.
    The experimental data is taken from Ref.~\onlinecite{Ishimaru:2002p055502}.
  }
  \end{center}
\end{figure}

Finally, we also report pair distribution functions of quenched amorphous silicon carbide.
Silicon-carbide is quenched at zero external pressure using the procedure outlined in the previous section for silicon.
Figure~\ref{g2} summarizes the results alongside experimental data from Ref.~\onlinecite{Ishimaru:2002p055502}.
All potentials reproduce the experimental pair distribution functions reasonably.
The unscreened potentials give pair distribution functions that are essentially indistinguishable from their screened counterparts and therefor not shown.
In all cases, the experimental data is broader than the data obtained from our simulations.
This is probably attributable to additional line broadening mechanisms that are active in the respective experimental setup.
Also, the experimental amorphous Si-C was created by ion irradiation and not by quenching, which could be the origin of some of the observed differences.

The notable differences between the two potentials are the heights of the nearest-neighbor peaks.
The TIII+S potential overestimates the height of the first neighbor peak significantly.
From the distribution functions for pure amorphous carbon (not shown) we see that this peak corresponds to the C-C bond length.
The TIII+S potential also appears to overestimate the peak at $2.5\,\text{\AA}$ that is barely visible in the EA+S simulation and the experimental data.
This length corresponds roughly to the Si-Si bond lengths.
Hence, the TIII+S appears to favor dimerization over the formation of a homogeneous melt, leading to a somewhat different structure than that found in experiments.

\section{Conclusions}

We have presented a simple method to augment existing bond-order potential by changing their cut-off procedure.
This fixes a number of issues with the description of non-equilibrium properties of matter, such as fracture or amorphous phase formation.
We here stress that \emph{without any reparameterization} of the potentials we are able to obtain correct cohesive stresses, proper bond-breaking in mode I cracks and appropriate properties of the amorphous phase.
Both the Tersoff III and Erhart \& Albe's potential are fitted to ground-state properties, yet they are able to reasonably describe these transition states.
The potential energy expression given by Eqs.~\eqref{bonden} to \eqref{angular1} is hence an exquisite extrapolation scheme.
Surely, this is due to the fact that there are good theoretical arguments~\cite{
Abell:1985p6184,
Horsfield:1996p12694,
Pettifor:1999p8487,
Finnis:2007p133} for this particular functional form.
Future work will focus on augmenting a recent potential for the ternary Si-C-H system in a similar manner.~\cite{Schall:2012}

Force routines for the potentials of this paper are available at the location given in Ref.~\onlinecite{code}.

\section*{Acknowledgements}
We thank Matous Mrovec for many fruitful discussions on interatomic potentials, Gianpietro Moras for carrying out the Stillinger-Weber simulations, Jan G. Korvink for pointing out the relationship between the screening approach and Voronoi tessellation, and George C. Abell for useful comments on the manuscript.
This work was supported by the German Federal Ministry of Education and Research (project ``OTRISKO''), the German Research Foundation (DFG Gu 367/30) and the European Commission (Marie-Curie IOF 272619 for L.P.).
Computations were carried out at Fraunhofer IWM and the J\"ulich Supercomputing Center.

\appendix


\end{document}